\documentclass{article}
\usepackage{amsmath}

\usepackage[english]{babel}
\usepackage[utf8]{inputenc}

\usepackage{amsfonts,epsfig}
\usepackage[hyphens]{url}
\usepackage{breakurl}

\usepackage{relsize}
\usepackage{fancyvrb}
\usepackage{amssymb}
\usepackage{geometry}
\usepackage{sectsty}
\usepackage{caption}
\usepackage{subcaption}
\usepackage[normalem]{ulem}
\sectionfont{\sffamily\bfseries\upshape\large}
\subsectionfont{\sffamily\bfseries\upshape\normalsize}
\subsubsectionfont{\sffamily\mdseries\upshape\normalsize}

\makeatletter
\renewcommand\@seccntformat[1]{\csname the#1\endcsname.\quad}
\makeatother

\renewcommand{\bibitem}{\vskip 2pt\par\hangindent\parindent\hskip-\parindent}

\makeatletter
\def\@maketitle{%
  \begin{center}%
  \let \footnote \thanks
    {\large \@title \par}%
    {\normalsize
      \begin{tabular}[t]{c}%
        \@author
      \end{tabular}\par}%
    {\small \@date}%
  \end{center}%
}
\makeatother

\title{\bf Many perspectives on Deborah Mayo's ``Statistical Inference as Severe Testing: How to Get Beyond the Statistics Wars''\footnote{We thank Deborah Mayo for writing the book that inspired this discussion, several blog commenters for valuable discussion, and the U.S. Office of Naval Research for partial support of this work through grant N000141912204.  Art Owen thanks Jessica Hwang for helpful comments and the National Science Foundation for grants IIS-1837931 and DMS-1521145.}\vspace{.1in}}
\author{Andrew Gelman, 
  Brian Haig,
   Christian Hennig,  
  Art B. Owen,  
  Robert Cousins,  
  Stan Young,  \\
  Christian Robert,
  Corey Yanofsky,
    E. J. Wagenmakers,  
  Ron Kenett, 
  Daniel Lakeland
  \vspace{.1in}}
\date{29 May 2019}
\vspace{.3in}

\begin{document}
\maketitle
\thispagestyle{empty}

\begin{abstract}
  The new book by philosopher Deborah Mayo is relevant to data science for topical reasons, as she takes various controversial positions regarding hypothesis testing and statistical practice, and also as an entry point to thinking about the philosophy of statistics.  The present article is a slightly expanded version of a series of informal reviews and comments on Mayo's book.  We hope this discussion will introduce people to Mayo's ideas along with other perspectives on the topics she addresses.  
\end{abstract}

\vspace{.05in}

\section{Background}

Deborah Mayo is a leading philosopher of science who, in her 1996 book, {\em Error and the Growth of Experimental Knowledge}, expanded Karl Popper's idea of falsification and Imre Lakatos's framework of scientific research programs so as to apply to statistical practice in real-world science.  In 2018 she released a new book, {\em Statistical Inference as Severe Testing: How to Get Beyond the Statistics Wars}, with the intention of addressing various controversies in statistics---ranging from longstanding disputes on Bayesian inference, likelihoods, and stopping rules, to more recent debates over the role of statistical methods in the replication crisis---from her neo-falsificationist perspective.

On the occasion of the release of Mayo's book, we contacted fifteen researchers in statistics and other sciences and gave them the opportunity to share reactions to {\em Statistical Inference as Severe Testing}.  Six people replied, with comments of length ranging from one paragraph to several pages.  We posted all these on our blog,\footnote{\url{https://statmodeling.stat.columbia.edu/2019/04/12/several-reviews-of-deborah-mayos-new-book-statistical-inference-as-severe-testing-how-to-get-beyond-the-statistics-wars/}} and then many others added their thoughts in the comment thread.

In the present article, we reproduce the invited reviews along with highlights from blog discussions.  Deborah Mayo participated in discussion as well, but rather than repeating her online contributions here, we hope that she will be able to share her reactions to all the comments in this article.

The point of this article is not to evaluate Mayo's book or to decide whether her ideas are correct.  Rather, we are using the publication of {\em Statistical Inference as Severe Testing}, and Mayo's vigorous advocacy of this material, as an opportunity to revisit some fundamental issues about statistical methods and philosophy.

\section{Review by Brian Haig}

We start with psychology researcher Brian Haig, because he's a strong supporter of Mayo's message and his review also serves as an introduction and summary of her ideas.  Haig asks what can psychology's statistics reformers learn from the error-statistical severe testing perspective.

\begin{quotation}

\subsection*{Introduction}
Deborah Mayo's ground-breaking book, {\em Error and the Growth of Statistical Knowledge} (1996), which won the Lakatos Prize for outstanding work in the philosophy of science, presented the first extensive formulation of her error-statistical perspective on statistical inference. Its novelty lay in the fact that it employed ideas in statistical science to shed light on philosophical problems to do with evidence and inference.

By contrast, Mayo's just-published book, {\em Statistical Inference as Severe Testing} (SIST) (2018), focuses on problems arising from statistical practice (``the statistics wars''), but endeavors to solve them by probing their foundations from the vantage points of philosophy of science, and philosophy of statistics. The ``statistics wars'' to which Mayo refers concern fundamental debates about the nature and foundations of statistical inference. These wars are longstanding and recurring. Today, they fuel the ongoing concern many sciences have with replication failures, questionable research practices, and the demand for an improvement of research integrity. By dealing with the vexed problems of current statistical practice, SIST is a valuable repository of ideas, insights, and solutions designed to help a broad readership deal with the current crisis in statistics.

Psychology has been prominent among a number of disciplines suggesting statistical reforms that are designed to improve replication and other research practices. Despite being at the forefront of these reforms, psychology has, I believe, ignored the philosophy of statistics at its cost. In this post, I want to briefly consider two major proposals in psychology's methodological literature: The recommendation that psychology should employ the so-called ``new statistics'' in its research practice, and the alternative proposal that it should embrace Bayesian statistics. I do this from the vantage point of the error-statistical severe testing perspective, which, for my money, is the most coherent philosophy of statistics currently available. I have enumerated what I take to be its strengths elsewhere (Haig, 2017). Of course, not everyone will want to endorse this perspective, but it surely needs to be taken seriously by methodologists and researchers, in psychology and further afield. Before concluding, I will also offer some remarks about two interesting features of the conception of science adopted in SIST, along with a brief comment about the value of the philosophy of statistics.

\subsection*{The new statistics}
For decades, numerous calls have been made for replacing tests of statistical significance with alternative statistical methods. The new statistics, a package deal comprising effect sizes, confidence intervals, and meta-analysis, is one reform movement that has been heavily promoted in psychological circles (Cumming, 2012, 2014) as a much needed successor to null hypothesis significance testing (NHST), which is deemed to be broken-backed. Eric Eich, the recent editor of Psychological Science, which is the flagship journal of the Association for Psychological Science, endorsed the use of the new statistics, wherever
appropriate. The new statistics might be considered the Association's current quasi-official position on statistical inference, despite the appearance of a number of public criticisms of the approach by both frequentists and Bayesians. It is noteworthy that the principal advocates of the new statistics have not publicly replied to these criticisms. Although SIST does not directly address the new statistics movement, its suggestions for overcoming the statistics wars contain insights about statistics that that can be directly employed to mount a powerful challenge to that movement.

\paragraph{Null hypothesis significance testing.} The new statisticians recommend replacing NHST with their favored statistical methods by asserting that it has several major flaws. Prominent among them are the familiar claims that NHST encourages dichotomous thinking, and that it comprises an indefensible amalgam of the Fisherian and Neyman-Pearson schools of thought. However, neither of these features applies to the error-statistical understanding of NHST. The claim that we should abandon NHST because it leads to dichotomous thinking is unconvincing because it is leveled at the misuse of a statistical test that arises from a poor understanding of its foundations. Fisher himself explicitly cautioned against such thinking. Further, SIST makes clear that the common understanding of the amalgam that is NHST is not an amalgam of Fisher's and Neyman and Pearson's thinking on the matter, especially their mature thought. Further, the error-statistical outlook can accommodate both evidential and behavioral interpretations of NHST, respectively serving probative and performance goals, to use Mayo's suggestive terms. SIST urges us to move beyond the claim that NHST is an inchoate hybrid. Based on a close reading of the historical record, Mayo argues that Fisher and Neyman and Pearson should be interpreted as compatibilists, and that focusing on the vitriolic exchanges between Fisher and Neyman prevents one from seeing how their views dovetail. Importantly, Mayo formulates the error-statistical perspective on NHST by assembling insights from these founding fathers, and additional sources, into a coherent hybrid.

Tellingly, the recommendation of the new statisticians to abandon NHST, understood as an inchoate hybrid, commits the fallacy of the false dichotomy because there exist alternative defensible accounts of NHST. The error-statistical perspective is one of these. The neo-Fisherian outlook of Hurlbert and Lombardi (2009) is another (Haig, 2017).

\paragraph{Confidence intervals.} For the new statisticians, confidence intervals replace p-valued null hypothesis significance testing. Confidence intervals are said to be more informative, and more easily understood, than p values, as well as serving the important scientific goal of estimation, which is preferred to hypothesis testing. Both of these claims are open to challenge. Whether confidence intervals are more informative than statistical hypothesis tests in a way that matters will depend on the research goals being pursued. For example, p values might properly be used to get a useful initial gauge of whether an experimental effect occurs in a particular study, before one runs further studies and reports p values, supplementary confidence intervals and effect sizes. The claim that confidence intervals are more easily understood than p values is surprising, and is not borne out by the empirical evidence (e.g., Hoekstra, Morey, Rouder, and Wagenmakers, 2014). I will speak to the claim about the greater importance of estimation under the next heading.

There is a double irony in the fact that the new statisticians criticize NHST for encouraging simplistic dichotomous thinking: As already noted, such thinking is straightforwardly avoided by employing tests of statistical significance properly, whether or not one adopts the error-statistical perspective. For another, the adoption of standard frequentist confidence intervals in place of NHST forces the new statisticians to engage in dichotomous thinking of another kind: A parameter estimate is either inside, or outside, its confidence interval.

Error-statisticians have good reason for claiming that their reinterpretation of frequentist confidence intervals is superior to the standard view. The standard account of confidence intervals adopted by the new statisticians prespecifies a single confidence interval (a strong preference for 0.95 in their case). The single interval estimate corresponding to this level provides the basis for the inference that is drawn about the parameter values, depending on whether they fall inside or outside the interval. A limitation of this way of proceeding is that each of the values of a parameter in the interval are taken to have the same probative force, even though many will have been weakly tested. By contrast, the error-statistician draws inferences about each of the obtained values according to whether they are warranted, or not, at different severity levels, thus leading to a series of confidence intervals. Crucially, the different values will not have the same probative force. Clearly, this is a more nuanced and informative assessment of parameter estimates than that offered by the standard view. Details on the error-statistical conception of confidence intervals can be found in SIST (pp.\ 189--201), as well as Mayo and Spanos (2011) and Spanos (2014).

Assuming that the new statisticians want to adopt a sound frequentist conception of confidence intervals, they would improve their practice by moving to the superior error-statistical understanding of such intervals.

\paragraph{Estimation and hypothesis tests.} The new statisticians claim, controversially, that parameter estimation, rather than statistical hypothesis testing, leads to better science. Their preference for estimation leads Cumming (2012) to aver that typical questions in science are what questions (e.g., ``What is the age of the earth?'', ``What is the most likely sea-level rise by 2012?''). Explanatory why questions and how questions, the latter which usually ask for information about causal mechanisms, are not explicitly considered. However, why and how questions are just as important as what questions for science. They are the sort of questions that science seeks to answer when constructing and evaluating hypotheses and theories. By contrast, SIST makes clear that, with its error-statistical perspective, statistical inference can be employed to deal with both estimation and hypothesis testing problems. It also endorses the view that providing explanations of things is an important part of science.

\subsection*{Bayesian statistics}

Unlike the field of statistics, the Bayesian outlook has taken an age to assert itself in psychology. However, a cadre of methodologists has recently advocated the use of Bayesian statistical methods as a superior alternative to the messy frequentist practice that dominates psychology's research landscape (e.g., Dienes, 2011, Wagenmakers, 2007). These Bayesians criticize NHST, often advocate the use of Bayes factors for hypothesis testing, and rehearse a number of other well-known Bayesian objections to frequentist statistical practice. Of course, there are challenges for Bayesians from SIST, just as there are for the new statisticians. They also need to reckon with the coherent hybrid NHST produced by the error statisticians, and argue against it if they want to justify abandoning NHST; they need to rethink whether Bayes factors can provide strong tests of hypotheses without their ability to probe errors; and they should consider, among other challenges, Mayo's critique of the likelihood principle, a principle to which they often appeal when critiquing frequentist statistics.

\paragraph{Contrasts with the error-statistical perspective.} One of the major achievements of SIST is that it provides a comprehensive critical evaluation of the major variants of Bayesian statistical thinking, including the default, pragmatic, and eclectic options within the Bayesian corpus. SIST contains many challenges for Bayesians to consider. Here, I want to note three basic features of Bayesian thinking, which are rejected by the error-statistical approach of SIST:

First, the error-statistical approach rejects the Bayesian insistence on characterizing the evidential relation between hypothesis and evidence in a universal and logical manner in terms of Bayes' theorem. By contrast, it formulates the relation in terms of the substantive and specific nature of the hypothesis and the evidence with regards to their origin, modeling, and analysis. This is a consequence of a strong commitment to a contextual approach to testing using the most appropriate frequentist methods available for the task at hand.

Second, the error-statistical philosophy also rejects the classical Bayesian commitment to the subjective nature of fathoming prior probabilities in favor of the more objective process of establishing error probabilities understood in frequentist terms. It also finds the turn to objective Bayesianism unsatisfactory, as SIST makes clear.

Third, the error-statistical outlook employs probabilities to measure how effectively methods facilitate the detection of error, and how those methods enable us to choose between alternative hypotheses. Orthodox Bayesians are not concerned with error probabilities at all. Instead, they use probabilities to measure belief in hypotheses or degrees of confirmation. It is for this reason that error-statisticians will say about Bayesian methods that, without supplementation with error probabilities, they are not capable of providing stringent tests of hypotheses.

\paragraph{The Bayesian remove from scientific practice.} Two additional features of the Bayesian focus on beliefs, which have been noted by philosophers of science, draw attention to their outlook on science. First, Kevin Kelly and Clark Glymour worry that ``Bayesian methods assign numbers to answers instead of producing answers outright.''\ (2004, p.\ 112) \ Mayo agrees that we should focus on the phenomena of interest, not the epiphenomena of degrees of belief. And second, Henry Kyburg is puzzled by the Bayesian's desire to ``replace the fabric of science \dots with a vastly more complicated representation in which each statement of science is accompanied by its probability, for each of us.''\ (1992, p.\ 149) \ Kyburg's puzzlement prompts the question, Why should we be interested in each other's probabilities? This is a question raised by David Cox about prior probabilities, and noted in SIST. I think that these legitimate expressions of concern stem from the reluctance of many Bayesians to study science itself. This Bayesian remove from science contrasts markedly with SIST's direct engagement with science. Mayo is a philosopher of science as well as statistics, and has a keen eye for scientific practice. Given that contemporary philosophers of science tend to take scientific practice seriously, it comes as no surprise that, in SIST, Mayo brings it to the fore when dealing with statistics. Indeed, her error-statistical philosophy should be seen as a significant contribution to the so-called new experimentalism, with its strong focus, not just on experimental practice in science, but also on the role of statistics in such practice. Her discussion of the place of frequentist statistics in the discovery of the Higgs boson in particle physics is an instructive case in point.
Taken together, these just-mentioned points of difference between the Bayesian and error-statistical philosophies constitute a major challenge to Bayesian thinking in psychology, and elsewhere, that methodologists, statisticians, and researchers need to consider.

\paragraph{Bayesianism with error-statistical foundations.} One particularly important modern variant of Bayesian thinking, which receives attention in SIST, is the falsificationist Bayesianism of Andrew Gelman, which received its major formulation in Gelman and Shalizi (2013). Interestingly, Gelman regards his Bayesian philosophy as essentially error-statistical in nature---an intriguing claim, given the anti-Bayesian preferences of both Mayo and Gelman's co-author, Cosma Shalizi. Gelman's philosophy of Bayesian statistics is also significantly influenced by Popper's view that scientific propositions are to be submitted to repeated criticism in the form of strong empirical tests. For Gelman, best Bayesian statistical practice involves formulating models using Bayesian statistical methods, and then checking them through hypothetico-deductive attempts to falsify and modify those models.
Both the error-statistical and neo-Popperian Bayesian philosophies of statistics extend and modify Popper's conception of the hypothetico-deductive method, while at the same time offering alternatives to received views of statistical inference. The error-statistical philosophy injects into the hypothetico-deductive method an account of statistical induction that employs a panoply of frequentist statistical methods to detect and control for errors. For its part, the Bayesian alternative involves formulating models using Bayesian statistical methods, and then checking them through attempts to falsify and modify those models. This differs from the received philosophy of Bayesian statistical modeling, which is regarded as a formal inductive process.

From the wide-ranging evaluation in SIST of the major varieties of Bayesian statistical thought on offer, Mayo concludes that Bayesian statistics needs new foundations: In short, those provided by her error-statistical perspective. Gelman acknowledges that his falsificationist Bayesian philosophy is underdeveloped, so it will be interesting to see how its further development relates to Mayo's error-statistical perspective. It will also be interesting to see if Bayesian thinkers in psychology engage with Gelman's brand of Bayesian thinking. Despite the appearance of his work in a prominent psychology journal, they have yet to do so. However, Borsboom and Haig (2013) and Haig (2018), provide sympathetic critical evaluations of Gelman's philosophy of statistics. Mayo's treatment of Gelman's philosophy brings to notice the interesting point that she is willing to allow a decoupling of statistical outlooks and their traditional philosophical foundations in favor of different foundations, which are judged more appropriate.

\subsection*{SIST and the nature of science}

Before concluding, I [Haig] want to convey some sense of how SIST extends our understanding of the nature of science. I restrict my attention to its treatment of the process of falsification and the structure of modeling, before saying something about the value of philosophy for statistics.

\paragraph{Falsificationism.} Probably the best known account of scientific method is Karl Popper's falsificationist construal of the hypothetico-deductive method, understood as a general strategy of conjecture and refutation. Although it has been roundly criticized by philosophers of science, it is frequently cited with approval by scientists, including psychologists, although they seldom consider it in depth. One of the most important features of SIST is its presentation of a falsificationist view of scientific inquiry, with error statistics serving an indispensable role. From a sympathetic, but critical, reading of Popper, Mayo endorses his strategy of developing scientific knowledge by identifying and correcting errors through strong tests of scientific claims. Making good on Popper's lack of knowledge of statistics,
Mayo shows how one can properly employ a range of, often familiar, error-statistical methods to implement her all-important severity requirement. Stated minimally, and informally, this requirement says ``A claim is severely tested to the extent that it has been subjected to and passes a test that probably would have found flaws, were they present.''\ (SIST, p.\ xii) \ Further, in marked contrast with Popper, who deemed deductive inference to be the only legitimate form of inference, Mayo's conception of falsification stresses the importance of inductive, or content-increasing, inference in science. We have here, then, a viable account of falsification, which goes well beyond Popper's account with its lack of operational detail about how to construct strong tests. It is worth noting that SIST presents the error-statistical stance as a constructive interpretation of Fisher's oft-cited remark that the null hypothesis is never proved, only possibly disproved.

\paragraph{A hierarchy of models.} Building on Patrick Suppes' (1962) insight that science employs a hierarchy of models that ranges from experimental experience to theory, Mayo's (1996) error-statistical philosophy adopts a framework in which three different types of models are interconnected and serve to structure error-statistical inquiry: Primary models, experimental models, and data models. Primary models break down a research question into a set of local hypotheses that can be investigated using reliable methods. Experimental models structure the particular models at hand, and serve to link primary models to data models. And, data models generate and model raw data, as well as checking whether the data satisfy the assumptions of the experimental models. It should be mentioned that the error-statistical approach has been extended to primary models and theories of a more global nature and, in SIST, also includes a consideration of experimental design and the analysis and generation of data.

Interestingly, this hierarchy of models employed in the error-statistical perspective exhibits a structure similar to the important three-fold distinction between data, empirical phenomena, and theory (Bogen and Woodward, 1988, Haig, 2014). This related pair of three-fold distinctions accords much better with scientific practice than the coarse-grained, data-theory/model distinction that is ubiquitous in scientific talk. As with error-statistical modeling, Bogen and Woodward show that local data analysis and statistical inference connect to substantive theories via intermediate claims about experimental models and empirical phenomena. The detection of phenomena, typically empirical regularities in psychology, makes heavy-duty use of statistical methods. In this regard, SIST strongly endorses Fisher's caution against taking a single small p value as an indicator of a genuine effect.

Although researchers and textbook writers in psychology correctly assume that rejection of the null hypothesis implies acceptance of the alternative hypothesis, they too often err in treating the alternative hypothesis as a research, or scientific, hypothesis rather than as a statistical hypothesis. Substantive knowledge of the domain in question is required in order to formulate a scientific hypothesis that corresponds to the alternative statistical hypothesis. SIST explicitly forbids concluding that statistical significance implies substantive significance. In addition, it counsels that one should knowingly moving back and forth between the two types of significance in respect of the hierarchy of models just mentioned.

\paragraph{The philosophy of statistics.} A heartening attitude that comes through in SIST is the firm belief that a philosophy of statistics is an important part of statistical thinking. This contrasts markedly with much of statistical theory, and most of statistical practice. Through precept and practice, Mayo's work makes clear that philosophy can have a direct impact on statistical practice. Given that statisticians operate with an implicit philosophy, whether they know it or not, it is better that they avail themselves of an explicitly thought-out philosophy that serves practice in useful ways. SIST provides a strong philosophical aid to an improved understanding and use of tests of statistical significance and other frequentist statistical methods. More than this, SIST is a book on scientific methodology in the proper sense of the term. Methodology is the interdisciplinary field that draws from disciplines that include statistics, philosophy of science, history of science, as well as indigenous contributions from the various substantive disciplines. As such, it is the key to a proper understanding of statistical and scientific methods. Mayo's latest book is deeply informed about the philosophy, history, and theory of statistics, as well as statistical practice. It is for this reason that it is able to position the reader to go beyond the statistics wars.

\subsection*{Conclusion}
SIST provides researchers and methodologists with a distinctive perspective on statistical inference. Mayo's Popper-inspired emphasis on strong tests is a welcome antidote to the widespread practice of weak hypothesis testing in psychological research that Paul Meehl labored in vain to correct. Psychologists can still learn much of value from Meehl about the nature of good statistical practice in science, but Mayo's SIST contains understandings and recommendations about sound statistical practices in science that will take them further, if they have a mind to do so. In this post, I have invited the new statisticians and Bayesians to address the challenges to their outlooks on statistics that the error-statistical perspective provides.

\end{quotation}

\noindent
I [Gelman] just have one quick reaction here.  In his discussion of different schools of statistics, Haig writes:
\begin{quotation}
  \noindent
  There is a double irony in the fact that the new statisticians criticize NHST for encouraging simplistic dichotomous thinking: As already noted, such thinking is straightforwardly avoided by employing tests of statistical significance properly, whether or not one adopts the error-statistical perspective. For another, the adoption of standard frequentist confidence intervals in place of NHST forces the new statisticians to engage in dichotomous thinking of another kind: A parameter estimate is either inside, or outside, its confidence interval.
\end{quotation}
I don't think that's quite right.  A confidence or interval (or simply an estimate with standard error) can be used to give a sense of inferential uncertainty. There is no reason for dichotomous thinking when confidence intervals, or uncertainty intervals, or standard errors, are used in practice.

\begin{figure}
  \centerline{\includegraphics[width=.5\textwidth]{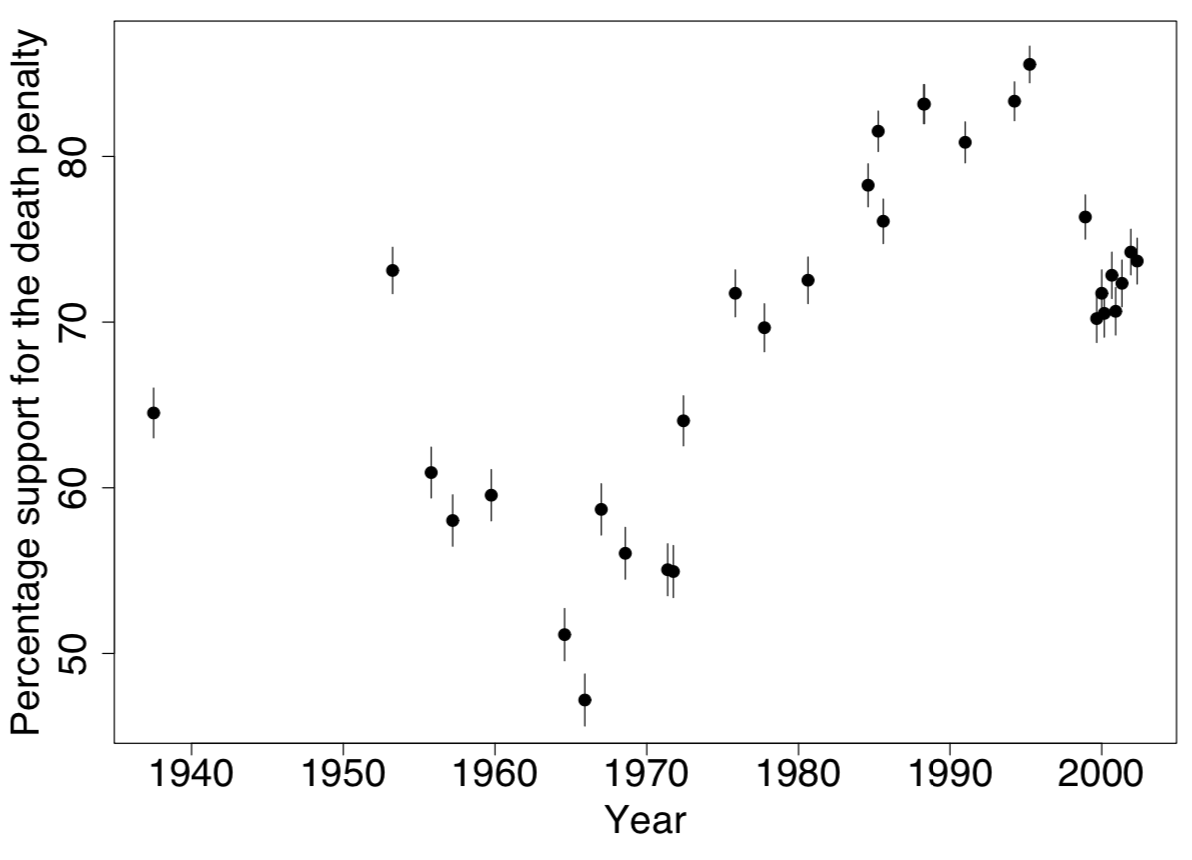}}
  \caption{\em Illustration of visual comparison of confidence intervals. Graph displays the proportion of respondents supporting the death penalty (estimates $\pm$ standard error, from Gallup polls over time.  The point of this example (from Gelman and Hill, 2007) is that it is possible to use confidence intervals and uncertainty statements as inferential tools, without dichotomization of summaries.}
  \label{conf_intervals}
  \end{figure}

  Figure \ref{conf_intervals} shows a simple example:  this graph has a bunch of 68\% confidence intervals, with no dichotomous thinking in sight. In contrast, testing some hypothesis of no change over time, or no change during some period of time, would make no substantive sense and would just be an invitation to add noise to our interpretation of these data.

One place I agree with Haig and Mayo is in his statement that ``the view that providing explanations of things is an important part of science.'' This is an issue that Guido Imbens and I wrestled with in the context of causal inference, where statistical theory and methods are all focused on testing or estimating the effects of specified treatments, but where as scientists we are often interested in the more challenging question of discovering causes or telling a causal story.  We came to the conclusion that Why questions can be interpreted as model checks, or, one might say, hypothesis tests---but tests of hypotheses of interest, not of straw-man null hypotheses.

And I agree very much with Haig's general point:  ``Given that statisticians operate with an implicit philosophy, whether they know it or not, it is better that they avail themselves of an explicitly thought-out philosophy that serves practice in useful ways.''  To paraphrase the baseball analyst Bill James, the alternative to good philosophy is not ``no philosophy,'' it's ``bad philosophy.'' I've spent too much time seeing Bayesians avoid checking their models out of a philosophical conviction that subjective priors cannot be empirically questioned, and too much time seeing non-Bayesians produce ridiculous estimates that could have been avoided by using available outside information (Gelman and Hennig, 2017). There's nothing so practical as good practice, but good philosophy can facilitate both the development and acceptance of better methods.

\section{Reaction of Stan Young}

Stan Young, a statistician who has worked in the pharmaceutical industry, wrote:

\begin{quotation}
  \noindent
I've been reading at the Mayo book and also pestering where I think poor statistical practice is going on. Usually the poor practice is by non-professionals and usually it is not intentionally malicious however self-serving. But I think it naive to think that education is all that is needed. Or some grand agreement among professional statisticians will end the problems.

There are science crooks and statistical crooks and there are no cops, or very few.

That is a long way of saying, this problem is not going to be solved in 30 days, or by one paper, or even by one book or by three books (Chambers, 2017, Harris, 2017, and Hubbard, 2015).

I think a more open-ended and longer dialog would be more useful with at least some attention to willful and intentional misuse of statistics.
\end{quotation}

\section{Review by Christian Hennig}
Hennig, a statistician and my collaborator on a paper on going beyond the overworked concepts of subjectivity and objectivity in statistics (Gelman and Hennig, 2017), first sent some general comments on what he likes about Mayo's new book:

\begin{quotation}
  \noindent
  Before I start to list what I like about ``Statistical Inference as Severe Testing.'' I should say that I don't agree with everything in the book. In particular, as a constructivist I am skeptical about the use of terms like ``objectivity,'' ``reality,'' and ``truth'' in the book, and I think that Mayo's own approach may not be able to deliver everything that people may come to believe it could, from reading the book (although Mayo could argue that overly high expectations could be avoided by reading carefully).

So now, what do I like about it?

\begin{enumerate}
\item I agree with the broad concept of severity and severe testing. In order to have evidence for a claim, it has to be tested in ways that would reject the claim with high probability if it indeed were false. I also think that it makes a lot of sense to start a philosophy of statistics and a critical discussion of statistical methods and reasoning from this requirement. Furthermore, throughout the book Mayo consistently argues from this position, which makes the different ``Excursions'' fit well together and add up to a consistent whole.

\item I get a lot out of the discussion of the philosophical background of scientific inquiry, of induction, probabilism, falsification, and corroboration, and their connection to statistical inference. I think that it makes sense to connect Popper's philosophy to significance tests in the way Mayo does (without necessarily claiming that this is the only possible way to do it), and I think that her arguments are broadly convincing at least if I take a realist perspective of science (which as a constructivist I can do temporarily while keeping the general reservation that this is about a specific construction of reality which I wouldn't grant absolute authority).

\item I think that Mayo does by and large a good job listing much of the criticism that has been raised in the literature against significance testing, and she deals with it well. Partly she criticizes bad uses of significance testing herself by referring to the severity requirement, but she also defends a well understood use in a more general philosophical framework of testing scientific theories and claims in a piecemeal manner. I find this largely convincing, conceding that there is a lot of detail and that I may find myself in agreement with the occasional objection against the odd one of her arguments.

\item The same holds for her comprehensive discussion of Bayesian/probabilist foundations in Excursion 6. I think that she elaborates issues and inconsistencies in the current use of Bayesian reasoning very well, maybe with the odd exception.

\item I am in full agreement with Mayo's position that when using probability modeling, it is important to be clear about the meaning of the computed probabilities. Agreement in numbers between different ``camps'' isn't worth anything if the numbers mean different things. A problem with some positions that are sold as ``pragmatic'' these days is that often not enough care is put into interpreting what the results mean, or even deciding in advance what kind of interpretation is desired.

\item As mentioned above, I'm rather skeptical about the concept of objectivity and about an all too realist interpretation of statistical models. I think that in Excursion 4 Mayo manages to explain in a clear manner what her claims of ``objectivity'' actually mean, and she also appreciates more clearly than before the limits of formal models and their distance to ``reality,'' including some valuable thoughts on what this means for model checking and arguments from models.

So overall it was a very good experience to read her book, and I think that it is a very valuable addition to the literature on foundations of statistics.
\end{enumerate}
\end{quotation}

Hennig also sent some specific discussion of  parts of Excursion 4 of Mayo (2018) titled Objectivity and Auditing, starting with section title ``The myth of ‘The myth of objectivity{'}'':

\begin{quotation}
  \noindent
  Mayo advertises objectivity in science as central and as achievable.
In contrast, in Gelman and Hennig (2017) we write: ``We argue that the words ‘objective' and ‘subjective' in statistics discourse are used in a mostly unhelpful way, and we propose to replace each of them with broader collections of attributes.'' I will here outline agreement and disagreement that I have with Mayo's Excursion 4, and raise some issues that I think require more research and discussion.

 \subsection*{Pushback and objectivity}

The second paragraph of Excursion 4 states in bold letters: ``The Key Is Getting Pushback,'' and this is the major source of agreement between Mayo's and my views. I call myself a constructivist, and this is about acknowledging the impact of human perception, action, and communication on our world-views, see Hennig (2010). However, it is an almost universal experience that we cannot construct our perceived reality as we wish, because we experience ``pushback'' from what we perceive as ``the world outside.'' Science is about allowing us to deal with this pushback in stable ways that are open to consensus. A major ingredient of such science is the ``Correspondence (of scientific claims) to observable reality,'' and in particular ``Clear conditions for reproduction, testing and falsification,'' listed as ``Virtue 4/4(b)'' in Gelman and Hennig (2017). Consequently, there is no disagreement with much of the views and arguments in Excursion 4 (and the rest of the book). I actually believe that there is no contradiction between constructivism understood in this way and Chang's (2012) ``active scientific realism'' that asks for action in order to find out about ``resistance from reality,'' or in other words, experimenting, experiencing and learning from error.

If what is called ``objectivity'' in Mayo's book were the generally agreed meaning of the term, I would probably not have a problem with it. However, there is a plethora of meanings of ``objectivity'' around, and on top of that the term is often used as a sales pitch by scientists in order to lend authority to findings or methods and often even to prevent them from being questioned. Philosophers understand that this is a problem but are mostly eager to claim the term anyway; I have attended conferences on philosophy of science and heard a good number of talks, some better, some worse, with messages of the kind ``objectivity as understood by XYZ doesn't work, but here is my own interpretation that fixes it.'' Calling frequentist probabilities ``objective'' because they refer to the outside world rather than epsitemic states, and calling a Bayesian approach ``objective'' because priors are chosen by general principles rather than personal beliefs are in isolation also legitimate meanings of ``objectivity,'' but these two and Mayo's and many others (see also the Appendix of Gelman and Hennig, 2017) differ. The use of ``objectivity'' in public and scientific discourse is a big muddle, and I don't think this will change as a consequence of Mayo's work. I prefer stating what we want to achieve more precisely using less loaded terms, which I think Mayo has achieved well not by calling her approach ``objective'' but rather by explaining in detail what she means by that.

 \subsection*{Trust in models?}

In the remainder, I will highlight some limitations of Mayo's ``objectivity'' that are mainly connected to Tour IV on objectivity, model checking and whether it makes sense to say that ``all models are false.'' Error control is central for Mayo's objectivity, and this relies on error probabilities derived from probability models. If we want to rely on these error probabilities, we need to trust the models, and, very appropriately, Mayo devotes Tour IV to this issue. She concedes that all models are false, but states that this is rather trivial, and what is really relevant when we use statistical models for learning from data is rather whether the models are adequate for the problem we want to solve. Furthermore, model assumptions can be tested and it is crucial to do so, which, as follows from what was stated before, does not mean to test whether they are really true but rather whether they are violated in ways that would destroy the adequacy of the model for the problem. So far I can agree. However, I see some difficulties that are not addressed in the book, and mostly not elsewhere either. Here is a list.

\paragraph{Adaptation of model checking to the problem of interest.}

As all models are false, it is not too difficult to find model assumptions that are violated but don't matter, or at least don't matter in most situations. The standard example would be the use of continuous distributions to approximate distributions of essentially discrete measurements. What does it mean to say that a violation of a model assumption doesn't matter? This is not so easy to specify, and not much about this can be found in Mayo's book or in the general literature. Surely it has to depend on what exactly the problem of interest is. A simple example would be to say that we are interested in statements about the mean of a discrete distribution, and then to show that estimation or tests of the mean are very little affected if a certain continuous approximation is used. This is reassuring, and certain other issues could be dealt with in this way, but one can ask harder questions. If we approximate a slightly skew distribution by a (unimodal) symmetric one, are we really interested in the mean, the median, or the mode, which for a symmetric distribution would be the same but for the skew distribution to be approximated would differ? Any frequentist distribution is an idealization, so do we first need to show that it is fine to approximate a discrete non-distribution by a discrete distribution before worrying whether the discrete distribution can be approximated by a continuous one? (And how could we show that?) And so on.

\paragraph{Severity of model misspecification tests.}

Following the logic of Mayo (2018), misspecification tests need to be severe in order to fulfill their purpose; otherwise data could pass a misspecification test that would be of little help ruling out problematic model deviations. I'm not sure whether there are any results of this kind, be it in Mayo's work or elsewhere. I imagine that if the alternative is parametric (for example testing independence against a standard time series model) severity can occasionally be computed easily, but for most model misspecification tests it will be a hard problem.

\paragraph{Identifiability issues, and ruling out models by other means than testing.}

Not all statistical models can be distinguished by data. For example, even with arbitrarily large amounts of data only lower bounds of the number of modes can be estimated; an assumption of unimodality can strictly not be tested (Donoho 1988). Worse, only regular but not general patterns of dependence can be distinguished from independence by data; any non-i.i.d. pattern can be explained by either dependence or non-identity of distributions, and telling these apart requires constraints on dependence and non-identity structures that can itself not be tested on the data (in the example given in 4.11 of Mayo, 2018, all tests discover specific regular alternatives to the model assumption). Given that this is so, the question arises on which grounds we can rule out irregular patterns (about the simplest and most silly one is ``observations depend in such a way that every observation determines the next one to be exactly what it was observed to be'') by other means than data inspection and testing. Such models are probably useless, however if they were true, they would destroy any attempt to find ``true'' or even approximately true error probabilities.

\paragraph{Robustness against what cannot be ruled out.}

The above implies that certain deviations from the model assumptions cannot be ruled out, and then one can ask: How robust is the substantial conclusion that is drawn from the data against models different from the nominal one, which could not be ruled out by misspecification testing, and how robust are error probabilities? The approaches of standard robust statistics probably have something to contribute in this respect (e.g., Hampel et al., 1986), although their starting point is usually different from ``what is left after misspecification testing.'' This will depend, as everything, on the formulation of the ``problem of interest,'' which needs to be defined not only in terms of the nominal parametric model but also in terms of the other models that could not be rules out.

\paragraph{The effect of preliminary model checking on model-based inference.}

Mayo is correctly concerned about biasing effects of model selection on inference. Deciding what model to use based on misspecification tests is some kind of model selection, so it may bias inference that is made in case of passing misspecification tests. One way of stating the problem is to realize that in most cases the assumed model conditionally on having passed a misspecification test does no longer hold. I have called this the ``goodness-of-fit paradox'' (Hennig, 2007); the issue has been mentioned elsewhere in the literature. Mayo has argued that this is not a problem, and this is in a well defined sense true (meaning that error probabilities derived from the nominal model are not affected by conditioning on passing a misspecification test) if misspecification tests are indeed ``independent of (or orthogonal to) the primary question at hand'' (Mayo 2018, p.\ 319). The problem is that for the vast majority of misspecification tests independence/orthogonality does not hold, at least not precisely. So the actual effect of misspecification testing on model-based inference is a matter that requires to be investigated on a case-by-case basis. Some work of this kind has been done or is currently done; results are not always positive (an early example is Easterling and Anderson, 1978).

 \subsection*{Conclusion}

The issues listed in Section 3 are in my view important and worthy of investigation. Such investigation has already been done to some extent, but there are many open problems. I believe that some of these can be solved, some are very hard, and some are impossible to solve or may lead to negative results (particularly connected to lack of identifiability). However, I don't think that these issues invalidate Mayo's approach and arguments; I expect at least the issues that cannot be solved to affect in one way or another any alternative approach. My case is just that methodology that is ``objective'' according to Mayo comes with limitations that may be incompatible with some other people's ideas of what ``objectivity'' should mean (in which sense it is in good company though), and that the falsity of models has some more cumbersome implications than Mayo's book could make the reader believe.
\end{quotation}

\section{Review by Art Owen}

Coming from a different perspective is theoretical statistician Art Owen.  I appreciate his efforts to link Mayo's words to the equations that we would ultimately need to implement, or evaluate, her ideas in statistics.

\begin{quotation}
  \noindent
For background on me, 
I think $p$-values are ok but one must understand their limitations.
They get treated as if a small value clinches an argument and ends
the discussion.  Instead, a small value means that a completely
null explanation is untenable, but it remains open whether some
specific alternative that the user has in mind is the right reason for 
rejecting the null.  The real reason might be 
a much desired causal outcome,
an unmeasured variable,
the multiplicity of testing,  
or some other model error such as correlations that were not
properly accounted for or even non-Gaussianity.
I liked Mayo's reminders that Fisher did not consider a single $p$-value
to be decisive.

More background: my favorite statistical tools are scatterplots and confidence intervals.

\subsection*{Power and severity}

There is an emphasis throughout on the importance of severe testing.
It has long been known that a test that fails to reject $H_0$ is not
very conclusive if it had low power to reject $H_0$.  So I wondered
whether there was anything more to the severity idea than that.
After some searching
I found on page 343 a description of how the severity idea differs from 
the power notion.

Suppose that one tests $H_0:\mu=\mu_0$ versus $H_1:\mu>\mu_0$ with
a one tailed test designed to have specific power in case that
$\mu=\mu_1$ for some $\mu_1>\mu_0$.
I believe that one tailed tests are more often than not a bad practice.
The justifications are usually that only one alternative direction is possible
or that only one alternative direction is consequential, and I think the
uncertainty around those statements will not be negligible compared to
the nominal level of the test.
Nevertheless, for purposes of discussion, let a one tailed test be based on random data $X$ and a statistic $d(X)$.
It rejects $H_0$ at level $\alpha$ if and only if $d(X)\ge c_\alpha$.

A post-hoc power analysis looks at $\mbox{Pr}( d(X)\ge c_\alpha;\mu_1)$.
A severity analysis looks at $\mbox{Pr}( d(X)\ge d(x);\mu_1)$ where
$x$ is the observed value of $X$.
If $H_0$ was not rejected then $d(x)<c_\alpha$ and so
\begin{equation}\label{eq:sevpow}
\mbox{Pr}(d(X)\ge d(x);\mu_1) \ge \mbox{Pr}(d(X)\ge c_\alpha;\mu_1).
\end{equation}
The severity number comes out higher than the power number. 

Mayo considers using power and severity as a post-hoc way to 
infer about whether $\mu<\mu_1$.
Here are those two choices followed by a third that I prefer:
\begin{itemize}
\item {\em Power analysis:} If $\mbox{Pr}( d(X)\ge c_\alpha;\mu_1)$ is high and 
$H_0$ was not rejected, it indicates evidence that $\mu\le\mu_1$.
\item {\em Severity analysis:} If $\mbox{Pr}( d(X)\ge d(x);\mu_1)$ is high and 
$H_0$ was not rejected, it indicates evidence that $\mu\le\mu_1$.
\item {\em Confidence  analysis:} If the confidence interval for $\mu$
is contained within $(-\infty,\mu_1]$, it indicates evidence that $\mu\le\mu_1$.
\end{itemize}
Equation~\ref{eq:sevpow} makes it clear that the severity analysis
will find evidence that $\mu\le\mu_1$ whenever the power analysis
does.

Let's double check the directionality of the tests and confidence interval. If a one tailed test rejects $H_0:\mu=\mu_0$
in favor of $H_0:\mu>\mu_0$ then it will ordinarily also reject
$H_0:\mu=\mu_{-1}$ too for any $\mu_{-1}<\mu_0$.  Then, inverting
our hypothetical one tailed test will give a one sided confidence interval from $-\infty$
up to some highest value, so it could well be a subset of $(-\infty,\mu_1]$.

I don't see any advantage to severity (or to posterior power) over
the confidence interval, if one is looking 
for evidence that $\mu\le \mu_1$.  One could replace the confidence interval 
by a posterior credible interval where that suits the problem and the user.

To make the case that severity is better than confidence,
it would be necessary to explain why a value of $\mu_1$
that is inside a confidence interval but fails a severity test should be considered
implausible, and similarly,  why a value of $\mu_1$ that lies outside of the confidence 
interval, should nonetheless be taken as plausible if it gets a low severity value.
If it can be proved that one of these outcomes is impossible then it would
be enough to explain why severity is better for the other one.

The idea of estimating power post-hoc has been criticized
as unnecessary.  I think that it might be useful in explaining a failure to reject $H_0$
as the sample size being too small.
A recent blog post by Andrew Gelman described how it is extremely hard to
measure power post hoc because there is too much uncertainty
about the effect size.  Then, even if you want it, you probably cannot reliably
get it.  I think severity is likely to be in the same boat.

\subsection*{One null and two alternative hypotheses, and multi-Bayes}

I liked the discussion of a remark attributed to Senn, that when $H_0:\mu=\mu_0$ is rejected
in favor of $H_A:\mu>\mu_0$ in a test with good power when $\mu\ge\mu_1$, this
is of course not evidence that $\mu\ge\mu_1$.  Nobody should think it was, 
and I have never encountered somebody who does, but
given how slippery it is to connect inferences to real world problems,
some people might.  
The usual description of $\mu_1$ is that of an effect so large that we would regret not detecting it. 
Rejecting $H_0$ lets us infer that $\mu>\mu_0$.  If any improvement over $\mu_0$ is enough
to justify some action, then rejecting $H_0$ gives us confidence of having done no harm
while the power calculation gives us some assurance that we won't miss a great benefit.
Things are more complicated if some values $\mu\in(\mu_0,\mu_1)$ are not really better
than $\mu_0$. For instance, the benefits expected by acting as if the alternative were
true could be proportional to $\mu-\mu_0$ minus some kind of switching cost.
Then $\mu$ has to be enough larger than $\mu_0$ to justify a change.
Meehl wrote about this point and maybe he was not the first.

One potential remedy is to construct a test so that with high
probability, the lower limit of the confidence interval is at least $\mu_1$.
This requires a new moving part: the value $\mu_2>\mu_1$ of $\mu$ at which to make this computation.
It cannot be $\mu_1$, because if $\mu=\mu_1$ it will not be feasible to get the lower limit to be above $\mu_1$
with high probability.  

One could make $\mu_1$ the new null. Then rejecting $H_0$ provides an inference
in favor of the proposed change being beneficial enough.
That still leaves open how to choose the second
alternative value $\mu_2>\mu_1$ under which the power is computed.

The problem seems to call for two different prior distributions.  The experimenter planning the
test might have a prior distribution on $\mu$ that is fairly optimistic about $\mu$
being meaningfully larger than $\mu_1$.  They might be aware of a more skeptical
prior distribution on $\mu$ that somebody else, who will judge their work, holds.  Then the idea is to
choose an experiment informative enough that the first person believes that
the second person will be convinced that
$\mu\in(\mu_1,\infty)$ has high probability.
That is we want
\begin{equation}\label{eq:multibayes}
\mbox{Pr}_{\rm experimenter}\, \mbox{Pr}_{\rm judge}( \mu>\mu_1\mid{\rm Data})
\ge 1-\epsilon_1)\ge1-\epsilon_2,
\end{equation}
for some small $\epsilon_j>0$.
The hard part of doing (\ref{eq:multibayes})  would be
in eliciting the two priors, learning first from the judge something about what
it would take to be convincing, and then pinning down the experimenter's beliefs.
There may also be the customary hard choice about what the variance of the future
data will be.  Picking the $\epsilon_j$ would also be tricky, not because of the
technical challenge, but just in eliciting loss functions.
Going through this put me in mind of the work of  Weerahandi  and Zidek (1981) on multi-Bayesian
theory.

It would be a nuisance if we had to consider whether the 
probabilistic beliefs of the judge and experimenter were subject to some
sort of statistical dependence.
This seems not to be the case.  Let the judge be convinced
if the Data belong to some set $S$.  Then we want the probability
under the experimenter's prior distribution that the data will belong to $S$.

\subsection*{Howlers and chestnuts}

I was intrigued by the collection of howlers and chestnuts.
Partisans of one statistical philosophy or
another might consider them one hit knockouts against another school of thought.
Maybe you can reduce an opposing school of thought to the butt of an xkcd joke 
and then not take it seriously any longer.  The problem is that the howlers have
to be constructed against a straw man.  For instance, a Bayesian howler against
frequentist methods could have a hypothetical situation with clear and obviously
important prior information ignored by the frequentist method.  That won't
generalize to cases with weaker and more ambiguous prior information.
Likewise a frequentist howler against Bayesian methods can be misleading.

More than one school of thought has had
contributions from excellent thinkers.  Choosing what to do cannot be as simple
as avoiding any approach with one or more howlers or chestnuts defined against it.
Statistical practice is full of tradeoffs and  catch-22s;
no method would remain if the howlers had their say.

\subsection*{Two numbers smushed into one}
Some attempts to fix $p$-value problems involve making the threshold
more stringent as $n$ increases.  I think this is a bad idea.  It is an attempt
to smush statistical and practical significance into one decision criterion.
Then nobody can undo the criterion to get back to statistical and practical
significance.  Presenting confidence intervals at one or more confidence levels is
better.  Then one can see the whole $2\times 2$ table:\\[3ex]
\begin{tabular}{lll}
& Statistically significant & Statistically insignificant \\
\hline
Practically significant & Interesting finding & We need more data\\
Practically insignificant &We could have used less data
& Maybe we can work with the null
\end{tabular}

\subsection*{Scientific and advocacy loss functions}
I believe that the statistical problem from incentives is more
severe than choice between Bayesian and frequentist methods 
or problems with people not learning how to use either kind of
method properly.  Incentive issues are more resistant to
education and naturally promote findings that don't reproduce.
This point has been made by many others.

We usually 
teach and do research assuming a scientific loss function that rewards being right. 
We have in mind a loss function like the scientist's loss function,

\begin{tabular}{ccc}
Scientist's loss& Decide A& Decide not A\\
A true &0 & 1 \\
A false & 1 & 0
\end{tabular}

\noindent
with generalizations to more than a binary choice and not necessarily equal losses. 
In practice many people using statistics are advocates.  They behave as if, or almost as if, the loss function is,

\begin{tabular}{ccc}
Advocate's loss & Decide A&Decide not A\\
A true &0 & 1 \\
A false & 0 & 1
\end{tabular}

\noindent
as it would be for one side in a civil lawsuit between two companies. 
The loss function strongly informs their analysis, be it Bayesian or
frequentist.  
The scientist and advocate both want to minimize their expected loss. 
They are lead to different methods. 

The issue can afflict scientists where A is a pet theory
(their own or their advisor's), people in business where A might be about
their product being safe enough to use or more effective than a competitor's,
and people working for social good, where A might be about results of
a past intervention or the costs and benefits of a proposed change.

If you're taking the scientific point of view but somehow hoping for
one outcome over the other, then your loss function starts to look like a 
convex combination of the above two loss functions.
You could then find yourself giving results you don't like extra
scrutiny compared to results that went your way.  This is similar
to confirmation bias, though that term is a better description of
biases from one's prior belief than from the loss function.

Conversely, even somebody working as an advocate may have a loss
function with a portion of scientific loss. For instance, advocating
for the scientifically wrong outcome too often will in some contexts
make one a less credible advocate.
\end{quotation}

\section{Reaction of Robert Cousins}

In some informal comments over email, physicist Robert Cousins addressed issues of statistical and practical significance:

\begin{quotation}
  \noindent
  Our [particle physicists'] problems and the way we approach them are quite different from some other fields of science, especially social science. As one example, I think I recall reading that you do not mind adding a parameter to your model, whereas adding (certain) parameters to our models means adding a new force of nature (!) and a Nobel Prize if true. As another example, a number of statistics papers talk about how silly it is to claim a $10^{-4}$ departure from 0.5 for a binomial parameter (ESP examples, etc), using it as a classic example of the difference between nominal (probably mismeasured) statistical significance and practical significance. In contrast, when I was a grad student, a famous experiment in our field measured a $10^{-4}$ departure from 0.5 with an uncertainty of 10\% of itself, i.e., with an uncertainty of $10^{-5}$. (Yes, the order of $10^{10}$ Bernoulli trials---counting electrons being scattered left or right.) This led quickly to a Nobel Prize for Steven Weinberg et al., whose model (now ``Standard'') had predicted the effect.
  \end{quotation}

I replied that
This interests me in part because I am a former physicist myself. I have done work in physics and in statistics, and I think the principles of statistics that I have applied to social science, also apply to physical sciences. Regarding the discussion of Bem's experiment, what I said was not that an effect of 0.0001 is unimportant, but rather that if you were to really believe Bem's claims, there could be effects of $+0.0001$ in some settings, $-0.0002$ in others, etc. If this is interesting, fine: I'm not a psychologist. One of the key mistakes of Bem and others like him is to suppose that, even if they happen to have discovered an effect in some scenario, there is no reason to suppose this represents some sort of universal truth. Humans differ from each other in a way that elementary particles to not.

To which Cousins replied:

\begin{quotation}
  \noindent
  Indeed in the binomial experiment I mentioned, controlling unknown systematic effects to the level of $10^{-5}$, so that what they were measuring (a constant of nature called the Weinberg angle, now called the weak mixing angle) was what they intended to measure, was a heroic effort by the experimentalists.
\end{quotation}

\section{Reaction of E. J. Wagenmakers}

I'll follow up with a very short review, or, should I say, reaction-in-place-of-a-review, from psychometrician E. J. Wagenmakers:

\begin{quotation}
  \noindent
  I cannot comment on the contents of this book, because doing so would require me to read it, and extensive prior knowledge suggests that I will violently disagree with almost every claim that is being made. In my opinion, the only long-term hope for vague concepts such as the ``severity'' of a test is to embed them within a rational (i.e., Bayesian) framework, but I suspect that this is not the route that the author wishes to pursue. Perhaps this book is comforting to those who have neither the time nor the desire to learn Bayesian inference, in a similar way that homeopathy provides comfort to patients with a serious medical condition.
\end{quotation}

This mini-review may seem to be too empty to be worth quoting, but I have included it here because we necessarily often do form impressions of new books or articles with a quick read or only a glance at a title or abstract.  It can be useful to share these reactions too, as long as we carefully state where our impressions are coming from, as Wagenmakers does here.

\section{Review by Christian Robert}

Statistician Christian Robert offers a more detailed Bayesian criticism on the tone as well as the content of Mayo's book:

\begin{quotation}
  \noindent
I sort of expected a different content when taking the subtitle, How to get beyond the Statistics Wars, at face value. But on the opposite the book is actually very severely attacking anything not in the line of the Cox-Mayo severe testing line. Mostly Bayesian approach(es) to the issue! For instance, Jim Berger's construct of his reconciliation between Fisher, Neyman, and Jeffreys is surgically deconstructed over five pages and exposed as a Bayesian ploy. Similarly, the warnings from Dennis Lindley and other Bayesians that the $p$-value attached with the Higgs boson experiment are not probabilities that the particle does not exist are met with ridicule. (Another go at Jim's Objective Bayes credentials is found in the squared myth of objectivity chapter. Maybe more strongly than against staunch subjectivists like Jay Kadane. And yet another go when criticizing the Berger and Sellke, 1987, lower bound results. Which even extends to Val Johnson's UMP-type Bayesian tests.)

\begin{quotation}
  \noindent
``Inference should provide posterior probabilities, final degrees of support, belief, probability (…) not provided by Bayes factors.'' (p.\ 443)
\end{quotation}

Another subtitle of the book could have been testing in Flatland given the limited scope of the models considered with one or at best two parameters and almost always a normal setting. I have no idea whatsoever how the severity principle would apply in more complex models, with e.g. numerous nuisance parameters. By sticking to the simplest possible models, the book can carry on with the optimality concepts of the early days, like sufficiency (p.\ 147) and  monotonicity and uniformly most powerful procedures, which only make sense in a tiny universe.

\begin{quotation}
  \noindent
``The estimate is really a hypothesis about the value of the parameter.  The same data warrant the hypothesis constructed!'' (p.\ 92)
\end{quotation}

There is an entire section on the lack of difference between confidence intervals and the dual acceptance regions, although the lack of unicity in defining either of them should come as a bother. Especially outside Flatland. Actually the following section, from p.\ 193 onward, reminds me of fiducial arguments, the more because Schweder and Hjort are cited there. (With a curve like Figure 3.3. operating like a cdf on the parameter $\mu$ but no dominating measure!)

\begin{quotation}
  \noindent
``The Fisher-Neyman dispute is pathological: there's no disinterring the truth of the matter (…) Fisher grew to renounce performance goals he himself had held when it was found that fiducial solutions disagreed with them.'' (p.\ 390)
\end{quotation}

Similarly the chapter on the ``myth of the `the myth of objectivity{'}'' (p.\ 1221) is mostly and predictably targeting Bayesian arguments. The dismissal of Frank Lad's arguments for subjectivity ends with a rather cheap shot that it ``may actually reflect their inability to do the math''  (p.\ 228). And the dismissal of loss function requirements in Ziliak and McCloskey is similarly terse, if reminding me of Aris Spanos' own arguments against decision theory. (And the arguments about the Jeffreys-Lindley paradox as well.)

\begin{quotation}
  \noindent
``It's not clear how much of the current Bayesian revolution is obviously Bayesian.'' (p.\ 405)
\end{quotation}

The section (Tour IV) on model uncertainty (or against ``all models are wrong'') is somewhat limited in that it is unclear what constitutes an adequate (if wrong) model. And calling for the CLT cavalry as backup (p.\ 299) is not particularly convincing.

It is not that everything is controversial in SIST  and I found agreement in many (isolated) statements. Especially in the early chapters. Another interesting point made in the book is to question whether or not the likelihood principle at all makes sense within a testing setting. When two models (rather than a point null hypothesis) are X-examined, it is a rare occurrence that the likelihood factorises any further than the invariance by permutation of iid observations. Which reminded me of our earlier warning on the dangers of running ABC for model choice based on (model specific) sufficient statistics. Plus a nice sprinkling of historical anecdotes, esp. about Neyman's life, from Poland, to Britain, to California, with some time in Paris to attend Borel's and Lebesgue's lectures. Which is used as a background for a play involving Bertrand, Borel, Neyman and (Egon) Pearson. I also enjoyed the sections on reuniting Neyman-Pearson with Fisher, while appreciating that Deborah Mayo wants to stay away from the ``minefields'' of fiducial inference. With, most interestingly, Neyman himself trying in 1956 to convince Fisher of the fallacy of the duality between frequentist and fiducial statements (p.\ 390). Wisely quoting Nancy Reid at BFF4 stating the unclear state of affair on confidence distributions. And the final pages reawakened an impression I had at an earlier stage of the book, namely that the ABC interpretation on Bayesian inference in Rubin (1984) could come closer to Deborah Mayo's quest for comparative inference (p.\ 441) than she thinks, in that producing parameters producing pseudo-observations agreeing with the actual observations is an ``ability to test accordance with a single model or hypothesis.''

\begin{quotation}
  \noindent
``Although most Bayesians these days disavow classic subjective Bayesian foundations, even the most hard-nosed, `we're not squishy' Bayesians retain the view that a prior distribution is an important if not the best way to bring in background information.'' (p.\ 413)
\end{quotation}
\end{quotation}

\section{Reaction of Corey Yanofsky}
Yanofsky, who works as a data scientist, comments on one of the more technical aspects of Mayo's argument relating to what she calls the severity of a hypothesis test:

\begin{quotation}
  \noindent
  Mayo argues that tests can be interpreted in terms of well and poorly indicated discrepancies from the null hypothesis. Well-warranted hypotheses about discrepancies from the null are (we are told) exactly those that have passed a severe test. (We're taking the model as given here because we're presuming that the model assumptions have passed their own severe tests.) When she translates this severity concept into math we end up with the severity (SEV) function, which seems to refute criticisms of statistical tests as inherently dichotomous and to unify testing and estimation.

To my knowledge Mayo has only ever demonstrated the SEV function for the normal model with wither known or unknown variance, and it just so happens that the SEV functions in these problems are numerically equal to the corresponding Bayesian posteriors under reference priors. Her examples got the concept across but it left me (a firm Bayesian) at a loss because these particular results are already intuitive from a Bayesian point of view. Thus to me it seemed {\em possible} but not thereby {\em demonstrated} that the severity concept really works as a philosophical foundation for frequentist statistics.

So I sought to drive a wedge between the SEV function and the Bayesian posterior so that I could figure out if SEV really made more sense than Bayes (or made sense at all). It turns out that a group-sequential design---that is, with interim looks at the data and stopping rules that could terminate data collection before the maximum sample size was observed---provided the ideal test bed. This is a bit sticky because it is not obvious how to construct the SEV function in such cases---there's more than one way to order outcomes as more or less ``extreme.'' There is one ordering---stagewise ordering---that is preferred in the literature; insofar as we can take Mayo as justifying current practice from a philosophical standpoint rather than doing original research in statistics, it seems reasonable to take stagewise ordering as canonical.

I discovered that I could create a design in which there was a non-negligible probability that the SEV function would fail to converge to a step function located at the true parameter value even as the amount of information available to estimate the parameter grew without bound. This also occurs with the other orderings I found in the literature; see Yanofsky (2019).

Mayo insists that selection effects due to early stopping must be taken into account but doesn't work out the consequences of early stopping for her notion that tests can be interpreted in terms of well or poorly warranted discrepancies from the null. I went and did it and I found that the consequence is that the SEV function can fail to be consistent---the irony is that it is precisely the complications brought on by the relevance of stopping rules to error probabilities that reveal the deficiencies of the argument.
\end{quotation}

\section{Reaction of Ron Kenett}

Kenett, an industrial statistician, shared his perspective on controversies in statistical analysis:

\begin{quotation}
  \noindent
Statistics is about information quality. If you deal with lab data, clinical trials, industrial problems or marketing data, it is always about the generation of information, and statistics should enable information quality.

In treating this issue we suggested a framework with eight dimensions (Kenett and Shmueli, 2014, 2016). From this perspective you look at the generalisation of findings (the sixth information quality dimension). Establishing causality is different from a statement about association.

As pointed out, Mayo is focused on the evaluation aspects of statistics. Another domain is the way findings are presented. A suggestion I made in the past was to use alternative representations. The trick is to state both what you claim to have found and also what you do not claim is being supported by your study design and analysis. An explicit listing of such alternatives could be a possible requirement in research paper in a section titled ``Generalization of findings.'' Some clinical researchers have already embraced this idea. An example of this form translational medicine is in  Kenett and Rubinstein (2019). Type S errors (Gelman and Tuerlinkcx, 2000) are actually speaking this language and permit to evaluate alternative representations with meaning equivalence.

So, I claim that the ``statistics war'' is engaged in just part of the playing ground. Other aspects, relevant to data analysis, should also be covered. The challenge is to close the gap between academia and practitioners who need methods that work beyond a publication goal measured by journal impact factors.

Integration is the third information quality dimension and the collection of review in this post demonstrate, yet again, that proper integration provides enhanced information quality. 

For me, the eight dimensions of information quality form a comprehensive top-down framework of what statistics is about, and it is beyond statistics wars.
\end{quotation}

\section{Reaction of Daniel Lakeland}
Lakeland, who has a background in mathematical modeling and civil engineering, offers a slightly different perspective:

\begin{quotation}
  \noindent
  To me, at the heart of statistics is not frequentism, or propositional logic, but decision making.
There are basically two parts of decision making that I can think of: decision making {\em about our models} and decision making {\em about how we construe reality}. On this point, I think I'm very much with Christian Hennig's constructivist views. It's our perception of reality that we have control over, and reality keeps smacking us when we perceive it incorrectly.

For me, the thing we want from science is models of how the world works which generalize sufficiently to make consistently useful predictions that can guide decisions.

We care about quantum physics for example because it lets us predict that putting together certain elements in a crystal will result in us being able to control the resistance of that crystal to electric current, or generate light of a certain color or detect light by moving charges around or stuff like that. I'm not just talking about what you might call economic utility, but also intellectual utility. Our knowledge of quantum mechanics lets people like Hawking come up with his radiation theory, which lets us make predictions about things in the vastness of space very far from ``money making.'' Similarly, knowledge of say psychology can let us come up with theories about how best to communicate warnings about dangers in industrial plants, or help people have better interpersonal relationships, or learn math concepts easier, or whatever.

When it comes to decision making about our models, logic rules. If parameter equals $q$, then prediction in vicinity of $y$ \dots If model predicts $y$ strongly, and $y$ isn't happening, then parameter not close to $q$. That sort of thing.

Within the Bayesian formalism this is what Bayes does: it downweights regions of parameter space that cause predictions that fail to coincide with real data. Of course, that's only {\em within the formalism}. The best the formalism can do is compress our parameter space down to get our model as close as possible to the manifold where the ``true'' model lies. But that true model never is reachable. It's like the denizens of Flatland trying to touch the sphere as it hovers over their plane. So in this sense, the notion of propositional logic and converging to ``true'' can't be the real foundation of Bayes. It's a useful model of Bayes, and it might be true in certain circumstances (like when you're testing a computer program you could maybe have a true value of a parameter) but it isn't general enough. Another way to say this is ``all models are wrong, but some are useful.'' If all models are wrong, at least most of the time, then logical truth of propositions isn't going to be a useful foundation.

Where I think the resolution of this comes from is in decision making outside Bayes. Bayesian formalism gives us in the limit of sufficient data collection, a small range of ``good'' parameters, that predict ``best'' according to the description of what kind of precision should be expected from prediction, which for Bayes is the probability of data given model.  But what we need to do with this is make a decision: do we stick with this model, or do we work harder to get another one. Whether to do that or not comes down to utilitarian concepts: How much does it matter to us that the model makes certain kinds of errors?

{\em One} way to evaluate that decision is in terms of frequency. If what we care about from a model is that it provides us with a good estimate of the frequency with which something occurs, then obviously frequency will enter into our evaluation of the model. But this is just one way in which to make a decision. We may instead ask what the sum of the monetary costs of the errors will be through time, or what the distribution of errors bigger than some threshold for mattering is, or a tradeoff between the cost of errors and the cost of further research and development required to eliminate them. If it takes a thousand years to improve our model a little bit, it may be time to just start using the model to improve lives today. That sort of thing.

So I see Andrew's interest in frequency evaluation of Bayesian models as one manifestation of this broader concept of model checking as {\em fitness for purpose}. Our goal in Bayesian reasoning isn't to get “subjective beliefs” or “the true value of the parameter” or frequently be correct, or whatever, it's to extract useful information from reality to help us make better decisions, and we get to define ``better'' in whatever way we want. Mayo doesn't get to dictate that frequency of making errors is All That Matters, and Ben Bernanke doesn't get to tell us that Total Dollar Amounts are all that matter, and Feynman doesn't get to tell us that the 36th decimal place in a QED prediction is all that matters.

And this is why ``The Statistics Wars'' will continue, because The Statistics Wars are secretly about human values, and different people value things differently.
\end{quotation}

\section{My own reactions}

I [Gelman] am still struggling with the key ideas of Mayo's book.  Struggling is a good thing here, I think!

First off, I appreciate that Mayo takes my own philosophical perspective seriously---I'm actually thrilled to be taken seriously, after years of dealing with a professional Bayesian establishment tied to naive (as I see it) philosophies of subjective or objective probabilities, and anti-Bayesians not willing to think seriously about these issues at all---and I don't think any of these philosophical issues are going to be resolved any time soon.  I say this because I'm so aware of the big Cantor-size hole in the corner of my own philosophy of statistical learning: In any real application, I will build a series of models, and as my collaborators and I think harder and become aware of more information, we check our models, find problems with them, and work to improve them.  As Box (1980) put it, we are always going back and forth between data gathering, modeling, inference, and model criticism.

In statistics---maybe in science more generally---philosophical paradoxes are sometimes resolved by technological advances.  Back when I was a student I remember all sorts of agonizing over the philosophical implications of exchangeability, but now that we can routinely fit varying-intercept, varying-slope models with nested and non-nested levels and (we've finally realized the importance of) informative priors on hierarchical variance parameters, a lot of the philosophical problems have dissolved; they've become surmountable technical problems.  (For example:  should we consider a group of schools, or states, or hospitals, as ``truly exchangeable''?  If not, there's information distinguishing them, and we can include such information as group-level predictors in our multilevel model.  Problem solved.)

Rapid technological progress resolves many problems in ways that were never anticipated. (Progress creates new problems too; that's another story.)  I'm not such an expert on deep learning and related methods for inference and prediction---but, again, I think these will change our perspective on statistical philosophy in various ways.

This is all to say that any philosophical perspective is time-bound.  On the other hand, I don't think that Popper/Kuhn/Lakatos will ever be forgotten:  this particular trinity of twentieth-century philosophy of science has forever left us in a different place than where we were, a hundred years ago.

To return to Mayo's larger message:  I agree with Hennig that Mayo is correct to place evaluation at the center of statistics.

I've thought a lot about this, in many years of teaching statistics to graduate students. In a class for first-year statistics Ph.D. students, you want to get down to the fundamentals.

What's the most fundamental thing in statistics? Experimental design?  No.  You can't really pick your design until you have some sense of how you will analyze the data.  (This is a principle of the great Raymond Smullyan:  To understand the past, we must first know the future.)  So is data analysis the most fundamental thing?  Maybe so, but what method of data analysis?  Last I heard, there are many schools.  {\em Bayesian} data analysis, perhaps?  Not so clear; what's the motivation for modeling everything probabilistically?  Sure, it's coherent---but so is some mental patient who thinks he's Napoleon and acts daily according to that belief.  We can back into a more fundamental, or statistical, justification of Bayesian inference and hierarchical modeling by first considering the principle of external validation of predictions, then showing (both empirically and theoretically) that a hierarchical Bayesian approach performs well based on this criterion---and then following up with the Jaynesian point that, when Bayesian inference fails to perform well, this recognition represents additional information that can and should be added to the model.  

To continue, it still seems to me that {\em the most foundational principles of statistics are frequentist}.  Not unbiasedness, not $p$-values, and not type 1 or type 2 errors, but frequency properties nevertheless. Statements about how well your procedure will perform in the future, conditional on some assumptions of stationarity and exchangeability (analogous to the assumption in physics that the laws of nature will be the same in the future as they've been in the past---or, if the laws of nature are changing, that they're not changing very fast!  We're in Cantor's corner again).

So, I want to separate the principle of frequency evaluation---the idea that frequency evaluation and criticism represents one of the three foundational principles of statistics (with the other two being mathematical modeling and the understanding of variation)---from specific statistical methods, whether they be methods that I like (Bayesian inference, estimates and standard errors, Fourier analysis, lasso, deep learning, etc.) or methods that I suspect have done more harm than good or, at the very least, have been taken too far (hypothesis tests, p-values, so-called exact tests, so-called inverse probability weighting, etc.).  We can be frequentists, use mathematical models to solve problems in statistical design and data analysis, and engage in model criticism, without making decisions based on type 1 error probabilities etc.

To say it another way, bringing in the title of the book under discussion:  I would not quite say that statistical inference {\em is} severe testing, but I do think that severe testing is a crucial part of statistics.  I see statistics as an unstable mixture of inference conditional on a model (``normal science'') and model checking (``scientific revolution'').  Severe testing is fundamental, in that prospect of revolution is a key contributor to the success of normal science.  We lean on our models in large part because they have been, and will continue to be, put to the test. And we choose our statistical methods in large part because, under certain assumptions, they have good frequency properties.

And now on to Mayo's subtitle. I don't think her, or my, philosophical perspective will get us ``beyond the statistics wars'' by itself---but perhaps it will ultimately move us in this direction, if practitioners and theorists alike can move beyond naive confirmationist reasoning (see Gelman, 2014) toward an embrace of variation and acceptance of uncertainty.

I'll summarize by expressing agreement with Mayo's perspective that frequency evaluation is fundamental, while disagreeing with her focus on various crude (from my perspective) ideas such as type 1 errors and $p$-values.  When it comes to statistical philosophy, I'd rather follow Laplace, Jaynes, and Box, rather than Neyman, Wald, and Savage.  I'd rather fit models and then use the predictions, estimates, and understanding obtained from these models to make decisions---rather than try to embed inference and decision into an uncomfortably shared framework.

\section*{References}

\noindent

\bibitem
  Bogen, J., and Woodward, J. (1988). Saving the phenomena. Philosophical Review, 97, 303--352.

\bibitem
Borsboom, D., and Haig, B. D. (2013). How to practice Bayesian statistics outside the Bayesian church: What philosophy for Bayesian statistical modelling? British Journal of Mathematical and Statistical Psychology, 66, 39--44.

\bibitem Box, G. E. P. (1980).  Sampling and Bayes inference in
scientific modelling and robustness (with discussion and rejoinder).  Journal of the Royal
  Statistical Society A, 143, 383--430.

\bibitem Chambers, C. (2017).  The Seven Deadly Sins of Psychology. Princeton University Press.

\bibitem Chang, H. (2012). Is Water H2O? Evidence, Realism and Pluralism. Dordrecht: Springer.
  
\bibitem
Cumming, G. (2012). Understanding the new statistics: Effect sizes, confidence intervals, and meta-analysis. New York: Routledge.

\bibitem
Cumming, G. (2014). The new statistics: Why and how. Psychological Science, 25, 7--29.

\bibitem
Dienes, Z. (2011). Bayesian versus orthodox statistics: Which side are you on? Perspectives
on Psychological Science, 6, 274--290.

\bibitem Donoho, D. (1988). One-sided inference about functionals of a density. Annals of Statistics, 16, 1390--1420.

\bibitem Easterling, R. G., and Anderson, H. E. (1978). The effect of preliminary normality goodness of fit tests on subsequent inference. Journal of Statistical Computation and Simulation, 8, 1--11.
  
\bibitem
  Gelman, A. (2014).  Confirmationist and falsificationist paradigms of science.   Statistical Modeling, Causal Inference, and Social Science blog, 5 Sept.  \url{https://statmodeling.stat.columbia.edu/2014/09/05/confirmationist-falsificationist-paradigms-science/}

\bibitem Gelman, A., and Hennig, C. (2017). Beyond subjective and objective in statistics (with discussion and rejoinder). Journal of the Royal Statistical Society A,  180, 967--1033. 

\bibitem Gelman, A., and Hill, J. (2007).  Data Analysis Using Regression and Multilevel/Hierarchical Models.  Cambridge University Press.

\bibitem Gelman, A., and Imbens, G. (2013).  Why ask why?
Forward causal inference and reverse causal questions.  Technical report, Department of Statistics, Columbia University.

\bibitem
Gelman, A., and Shalizi, C. R. (2013). Philosophy and the practice of Bayesian statistics.
British Journal of Mathematical and Statistical Psychology, 66, 8--38.

\bibitem
Haig, B. D. (2014). Investigating the psychological world: Scientific method in the
behavioral sciences. Cambridge, Mass.: MIT Press.

\bibitem
Haig, B. D. (2017). Tests of statistical significance made sound. Educational and
Psychological Measurement, 77, 489--506.

\bibitem
Haig. B. D. (2018). The philosophy of quantitative methods. New York: Oxford
University Press.

\bibitem
  Hampel, F. R., Ronchetti, E. M., Rousseeuw, P. J., and Stahel, W. A. (1986). Robust Statistics. New York: Wiley.

\bibitem Harris, R. (2017). Rigor Mortis: How Sloppy Science Creates Worthless Cures, Crushes Hope, and Wastes Billions. New York: Basic Books.

\bibitem Hennig, C. (2007). Falsification of propensity models by statistical tests and the goodness-of-fit paradox. Philosophia Mathematica, 15, 166-192.

 \bibitem Hennig, C. (2010). Mathematical models and reality: a constructivist perspective. Foundations of Science, 15, 29–48.

\bibitem
Hoekstra, R., Morey, R. D., Rouder, J. N., and Wagenmakers, E. J. (2014). Robust
misinterpretation of confidence intervals. Psychonomic Bulletin and Review, 21, 1157--1164.

\bibitem Hubbard, R. (2015). Corrupt Research. London: Sage Publications.

\bibitem
Hurlbert, S. H., and Lombardi, C. M. (2009). Final collapse of the Neyman-Pearson decision
theoretic framework and rise of the neoFisherian. Annales Zoologici Fennici, 46, 311--349.

\bibitem
Kelly, K.T., and Glymour, C. (2004). Why probability does not capture the logic of scientific justification. In C. Hitchcock (Ed.), Contemporary debates in philosophy of science,
94--114. Malden, Mass: Blackwell.

\bibitem Kenett, R. S., and Shmueli, G. (2014).  On information quality (with discussion and rejoinder).  Journal of the Royal
  Statistical Society A, 177, 3--38.

\bibitem Kenett, R. S., and Shmueli, G. (2016). Information Quality: The Potential of Data and Analytics to Generate Knowledge.  New York:  Wiley.

  \bibitem Kenett, R. S., and Rubinstein, A. (2019).  Generalizing research findings for enhanced reproducibility: A translational medicine case study.  \url{https://papers.ssrn.com/sol3/papers.cfm?abstract_id=3035070}

\bibitem
Kyburg, H. (1992). The scope of Bayesian reasoning. In PSA: Proceedings of the biennial
meeting of the Philosophy of Science Association, Vol.\ 2, 139--152.
University of Chicago Press.

\bibitem
Mayo, D. G (1996). Error and the growth of experimental knowledge. 
University of Chicago Press.

\bibitem
Mayo, D. G. (2018). Statistical inference as severe testing: How to get beyond the statistics
wars. New York: Cambridge University Press.

\bibitem
Mayo, D. G., and Spanos, A. (2011). Error statistics. In P. S. Bandyopadhyay and M. R. Forster
(Eds.), Handbook of philosophy of science: Vol.\ 7. Philosophy of statistics, 153--198.
Amsterdam: Elsevier.

\bibitem
Spanos, A. (2014). Recurring controversies about P values and confidence intervals revisited.
Ecology, 95, 645--651.

\bibitem
Suppes, P. (1969). Models of data. In E. Nagel, P. Suppes, and A. Tarski (Eds.), Logic,
methodology, and philosophy of science: Proceedings of the 1960 International Congress, 252--261. Stanford University Press.

\bibitem
Wagenmakers, E. J. (2007). A practical solution to the pervasive problems of p values.
Psychonomic Bulletin and Review, 14, 779--804.

\bibitem Weerahandi, S., and Zidek, J. V. (1981).  Multi-Bayesian statistical decision theory.  Journal of the Royal
  Statistical Society A, 144, 85--93.

\bibitem Yanofsky, C. (2019).  The SEV function just plain doesn't work.  It's Chancy blog, 5 Feb.  \url{https://itschancy.wordpress.com/2019/02/05/the-sev-function-just-plain-doesnt-work/}

\end{document}